# Absence of Morphotropic Phase Boundary Effects in BiFeO$_3$-PbTiO$_3$ Thin Films Grown via a Chemical Multilayer Deposition Method

Ashish Garg[1*], Shashaank Gupta[1], Shuvrajyoti Bhattacharjee[2], Dhananjai Pandey[2], Vipul Bansal[3], Suresh K Bhargava[3], Ju Lin Peng[3]

[1]Department of Materials Science and Engineering, Indian Institute of Technology, Kanpur 208016, India
[2]School of Materials Science and Technology, Institute of Technology, Banaras Hindu University, Varanasi 221005, India
[3]School of Applied Sciences, RMIT University, Melbourne 3001, Australia

## Abstract

Here, we report the unusual behaviour shown by the (BiFeO$_3$)$_{1-x}$-(PbTiO$_3$)$_x$ (BF-$x$PT) films prepared using a multilayer deposition approach by chemical solution deposition method. Thin film samples of various compositions were prepared by depositing several bilayers of BF and PT precursors by varying the BF or PT layer thicknesses. X-ray diffraction showed that final samples of all compositions show mixing of the two compounds resulting in a single phase mixture, also confirmed by transmission electron microscopy. In contrast to bulk equilibrium compositions, our samples show a monoclinic (M$_A$ type) structure suggesting disappearance of morphotropic phase boundary (MPB) about $x$ = 0.30 as observed in the bulk. This is accompanied by the lack of any enhancement of remnant polarization at MPB as shown by the ferroelectric measurements. Magnetic measurements show that the magnetization of the samples increases with increasing BF content. Significant magnetization of the samples indicates melting of spin spirals in the BF-xPT arising from random distribution of iron atoms across the film. Absence of Fe$^{2+}$ ions in the films was corroborated by X-ray photoelectron spectroscopy measurements. The results illustrate that used thin film processing methodology significantly changes the structural evolution in contrast to predictions from the equilibrium phase diagram as well as modify the functional characteristics of BP-xPT system dramatically.

---

[*] Corresponding author, Email: ashishg@iitk.ac.in

**I. INTRODUCTION**

Recent years have witnessed tremendous research on multiferroic $BiFeO_3$ (BF) which shows co-existence of ferroelectric and magnetic ordering in the same phase (Wang *et al.*, 2003). BF makes continuous solid solution with many insulating $ABO_3$ structured perovskite oxides, such as $PbTiO_3$ (PT) (Fedulov *et al.*, 1964) and $BaTiO_3$ (Kumar *et al.*, 1998). In particular, the mixed $(1-x)BiFeO_3–xPbTiO_3$ (BF-xPT) system has generated significant interest because this system has been suggested to exhibit high piezoelectric coefficient and higher Curie temperature in comparison to conventionally used $Pb(Zr,Ti)O_3$ (Fedulov *et al.*, 1964). These two compounds form a continuous solid solution whose phase diagram exhibits a morphotropic phase boundary (MPB) (Fedulov *et al.*, 1964). The narrow morphotropic phase region (composition width $\Delta x \approx 0.04$) is characterized by co-existence of both tetragonal (T) and monoclinic (M) phases while outside the MPB region, only the T and M phases are observed for $x \geq 0.31$ and $x \leq 0.27$, respectively (Bhattacharjee *et al.*, 2007). BF-xPT also shows a high ferroelectric Curie temperature in the range 763 to 1100K (Kaczmarek et al., 1975), and displays an unusually large tetragonality (Sunder *et al.*, 1995; Bhattacharjee *et al.*, 2007), three times more than that of $PbTiO_3$. These unusual features in bulk have led to number of studies on BF-xPT solid solution thin films in the past few years, (Sakamoto *et al.*, 2006; Liu *et al.*, 2006; Khan *et al.*, 2007) typically reporting good quality hysteresis loop with high remnant polarization (Khan *et al.*, 2007) at temperatures below room temperature and at high frequencies (mostly above 1 kHz).

Recently we demonstrated excellent ferroelectric properties with high insulation resistance in BF-0.60PT films prepared using a multilayer chemical solution deposition method (Gupta *et al.*, 2009). The films also showed a change in the crystal structure of films from equilibrium tetragonal structure to monoclinic $M_A$ type structure. In this paper, we show how compositional changes in multilayer processed BF-xPT thin films result in unusual characteristics where we do not observe a MPB like behaviour across $x = 0.30$ either in structure or properties, elucidating the significance of processing methodology resulting in marked effect over structure evolution and properties of BF-xPT system.

## II. EXPERIMENTAL DETAILS

The films were prepared by chemical solution deposition method using a novel multilayer approach described in our previous work (Gupta *et al.*, 2009). BF and PT solutions were spun coated alternatively at a speed of 4000 rpm for 30 seconds on the platinized Si substrates to grow the alternating PT and BF layers. Each coating step was followed by drying for 12 minutes at 360°C. Finally after deposition of 16 bi-layers of BF and PT, whole stack was annealed at 700°C for 1 h in nitrogen atmosphere. Final samples were between 300-500 nm thick consisting of 16 bi-layers of BF and PT of different thicknesses to yield overall compositions of $x = 0.27, 0.33, 0.60, 0.70$ and $0.80$. For all the film compositions, PT layer was deposited before BF layers.

Thermoelectron ARL X'tra high resolution X-Ray Diffractometer was used for collecting powder X-ray diffraction (XRD) data. The package Fullprof (J. Rodriguez-Carvajal, FULLPROF, Laboratory Leon Brillouin _CEA-CNRS_CEA/Saclay, 91191 Gif sur Yvette Cedex, France, 2006) was used for Le-Bail refinements using the XRD data in the 2θ range of 20° to 100°. In the refinements, the data in the 2θ range of 36° to 43.3° and 85° to 88° were excluded due to strong texture effects and the substrate peaks. Further, coexistence of Pt (FCC) substrate peaks was also taken into account in the refinements.

Transmission electron microscope (Jeol 1010) was used for cross sectional imaging and compositional studies. Magnetic measurements were carried out using a Vibrating sample magnetometer (ADE-DMS EV-7VSM). The ferroelectric and leakage properties of the films were measured by Radiant Premier II ferroelectric tester. 200 µm diameter platinum electrodes were sputter deposited on the films to facilitate the electrical measurements. X-ray photoelectron spectroscopy (XPS) measurements on the films were carried out on a VG MicroTech ESCA 310F instrument at a vacuum better than $1 \times 10^{-9}$ Torr. The general scan and core level spectra for Fe2p, Bi4f and Pb4f were recorded with un-monochromatized Mg K$\alpha$ radiation (photon energy = 1253.6 eV) at a pass energy of 20 eV and electron takeoff angle of 90°. The core level binding energies (BEs) were aligned with the adventitious C 1s binding energy of 285 eV.

## III. RESULTS AND DISCUSSION

### A. Structural Characterization

Crystal structure and phase analysis of the samples was carried out by X-Ray diffraction studies conducted at room temperature. Figure 1 (a) shows the XRD patterns of all the samples. When the peak positions of our samples are compared with the standard ICCD XRD data for pure PT and BF, the structure does not fit to either the rhombohedral BF or the tetragonal PT phase. This suggests the intermixing of the PT and BF precursor layers during post-deposition heat treatment and formation of a solid-solution. No significant change in the peak profiles and their positions is observed even when the PT content of the BF-xPT thin film samples changes from $x = 0.27$ to $x = 0.80$ further affirming that the XRD peaks in Fig. 1(a) belong to a single solid-solution phase only in agreement with our previous finding on BF-0.60PT films (Gupta *et al.*, 2009). For the BF-xPT solid solutions in bulk, the structure is known to be tetragonal for more than 30 mole % PT and monoclinic for less than 28 mol% PT (Bhattacharjee *et al.*, 2007).

Interestingly and unexpectedly, the thin film multilayer samples do not show any obvious change of structure across the so-called bulk MPB composition. (Fedulov *et al.*, 1964)

To verify the compositional uniformity and mixing, we carried out TEM and EDAX measurements, Figure 1 (b) shows a transmission electron microscope image of a BF-xPT sample for $x = 0.27$. The image does not show any individual layer of BF or PT and the structure appears like a homogenous film with presence of nano-crystalline grains (depicted as layer 3). This again confirms the mixing of the two types of layers. EDAX measurements on the cross sectional SEM images of all samples verified the overall compositions of the samples within an error of ~5-10% for various elements.

Upon a closer examination of the XRD patterns shown in Fig 1(a), we find that there is a change in the relative intensities of the peaks in the 2θ range of 45-48° and 55-60° as the composition of the films changes. It is seen that as the PT content (x) increases, the peak in the 45-48° range becomes stronger and the peak between 55-60° gets weaker. Same is also true of the two peaks between 2θ range of 25° and 35°. Although, this change in the relative intensities of the two peaks is indicative of some structural change that may occur upon the change in the composition, it can also be associated with slight change in grain orientation in the thin film samples or due to thin film strain effects arising from the mismatch at the substrate- sample interface. We then analyzed our XRD data using LeBail refinement technique to capture the significance of this change in the relative intensities, using the procedure similar to that described in our previous work (Gupta *et al.*, 2009) where we determined the crystal symmetry of the samples containing upto 60% PT to be $M_A$ type (Vanderbilt and Cohen notation) (Vanderbilt and Cohen, 2001) monoclinic structure with Cm space group symmetry. The refinements, as shown in Fig. 2, yield slightly lower $\chi^2$ values, only by 0.21, for a tetragonal structure with space group P4mm for compositions with PT higher than 60%, not sufficient enough to explicitly justify a structural change to a tetragonal structure after $x = 0.60$. Moreover, at the outset, it is tempting to associate the presence of monoclinic structure in samples with PT contents up to ~ 60% and possibly a subtle change to tetragonal beyond this composition to a shift in the MPB composition of BF-xPT thin films, prepared by multilayer method, which otherwise occurs at ~30% PT in bulk solid solutions. However, again, notwithstanding the fact the subtle structure change as predicted by the Le-Bail refinement after $x = 0.60$ itself is not conclusive; we do not associate this with the MPB due to reasons mentioned above.

Further, the multilayer deposition technique may result in very fine compositional gradients across the sample thickness and may also cause strains in the films. Large full-width half maximum (FWHM) of the film peaks, higher than 0.5º, is also indicative of the compositional variations across the film thickness. Such a behaviour is supported by previously observed strain induced structural changes away from bulk structure in epitaxial BF films (Wang *et al.*, 2003) as well as in PT films (Catalan *et al.*, 2006). However structural deviations away from the equilibrium structure in solid-solutions showing MPB effect have remained unreported thus far and can only be attributed to films synthesis resulting in a single phase yet chemically inhomogeneous structure.

## B. Electrical Measurements

We carried out ferroelectric measurements on our samples at 200K. The measurements were made using a bipolar pulse at a frequency of 1 kHz. The measurement results, as shown in Figure 3, confirm the ferroelectric nature of both BF rich and PT rich samples. The remnant polarization of the BF-xPT sample with x = 0.30 was ~19.5 $\mu C/cm^2$ which is only slightly larger than the values obtained for PT rich samples but much smaller than the value reported for laser ablated

BF-xPT films of compositions close to the bulk MPB (Khan *et al.*, 2007). Importantly, the samples could withstand fields at least up to ~600 kV/cm, high enough for device applications, shows that samples have good insulation resistance. Such high insulation resistance in these samples can be attributed to the alteration in the film's crystal structure to $M_A$ type monoclinic structure as compared to equilibrium (bulk) structure. Interestingly, as expected in bulk compositions at $x$ = 0.30 i.e. MPB, any extraordinary enhancement in the ferroelectric polarization is not observed. This is a very interesting as well as important observation because on one hand single phase formation occurs as suggested by the XRD patterns and the TEM image, there appears to be incomplete mixing of atoms at atomic level unexpectedly resulting in disappearance of MPB due to peculiar processing technique.

Room temperature leakage characteristics (log J ($A/cm^2$) *vs* E (kV/cm)) of the samples are shown Figure 4. The figure shows that the smallest leakage current of the order of $10^{-6}$ $A/cm^2$ at 100 kV/cm is shown by samples containing maximum PT content *i.e.* $x$ = 0.80, indicating highly resistive nature of the samples, also supported by the ferroelectric hysteresis data. The leakage current increases as the BF content of the films increases and follows an expected trend because leakage resistance of PT is far superior to BF.

From the above electrical data, one can observe that there is no obvious enhancement in the ferroelectric polarization near the usual (bulk) MPB composition range around $x$ = 0.30. This is most likely due to the nature of film processing resulting in mixing of BF and PT to an extent that one observes a single phase XRD spectra and homogenous TEM micrographs, but possibly not sufficient enough to exhibit the MPB effect which is observed in chemically homogeneous BF-PT solid solution ceramics and thin films (Bhattacharjee *et al.*, 2007; Khan *et al.*, 2007). The results illustrate that processing methods can lead to noticeable changes in the structure and properties of the BF-PT thin films to the extent that hitherto observed MPB effects are not noticed.

**C. Magnetic Studies**

Magnetic measurements were conducted on all samples and the results are shown in Figure 5. The figure shows that all the samples possess a saturated magnetic hysteresis loop along with finite remnant magnetization. The highest magnetization of ~15 emu/cc is exhibited by BF-xPT sample with $x = 0.27$. We find that the saturation magnetization keeps increasing as the PT content of the films decreases until $x = 0.27$. This is in contrast to single crystal $BiFeO_3$ (Lebeugle *et al.*, 2007) which do not exhibit magnetic hysteresis loop. Although the magnetization values we observe are not very large and are of the order of 0.01 $\mu_B$ (Bohr magneton), these are still significant when compared to the values shown by pure BFO films. (Eerenstein *et al.*, 2005)

Previous studies suggest that finite magnetization in BF can be achievable by various means such as by melting the spiral spin cycloid via application of large enough magnetic fields (Kadomtseva *et al.*, 2004), or by substrate induced constraints in epitaxial $BiFeO_3$ thin films (Wang *et al.*, 2003), or by means of chemical substitutions (Fedulov *et al.*, 1964; Gabbasova *et al.*, 1991). In the present study, high magnetic moment could result either due to the thin film clamping effect or due to the presence of PT in the solid solution destroying the spin cycloid or structural change or a combination of these. Melting of spin cycloid is expected since the thickness of each layer is less than the length of spin cycloid of 62 nm. Moreover any inhomogenity in the composition would further enhance the breaking of such a spin cycloid resulting in a finite magnetization.

In order to verify if the weak ferromagnetism is due to the presence of mixed valence of Fe ions i.e. $Fe^{3+}$ and $Fe^{2+}$, we carried out the X-ray photo-electron spectroscopic measurements on our samples. Figure 6 shows the XPS spectra of Fe, Bi and Pb ions in two samples, $x = 0.33$ and 0.80, of two extreme compositions. Figure 6(a) shows the Fe $2p$ peaks for both the samples. There are two peaks present at 711.1 eV and 725.3 eV which correspond to Fe $2p_{3/2}$ and Fe $2p_{1/2}$ respectively and are ascribed to $Fe2p_{3/2}$-O and $Fe2p_{1/2}$-O bonds for $Fe^{3+}$. The satellite peak belonging to $Fe^{3+}$ appears at 718.8 eV. The expected value of major peak for $Fe^{2+}$ is 709.5 eV while its satellite peak is expected to occur at 716 eV (Schedel-Niedrig *et al.*, 1995). The absence of these two peaks is indicative of the absence of $Fe^{2+}$ in our films as also shown previously (Eerenstein *et al.*, 2005). Figure 6 (b) shows the Bi and Pb $4f_{5/2}$ and $4f_{7/2}$ doublets for the two films. First it is evident that the intensities of Bi $4f$ doublets as compared to Pb $4f$ doublets goes down significantly as the PT content of the films increased reflecting the change in composition. One may also notice that in the sample with larger PT content, the peaks are slightly shifted towards higher binding energies suggesting the influence of cation substitution and hence altered bond characteristics.

The presence of magnetism in $BiFeO_3$ thin films has been a subjected of intense debate. For instance, while Wang *et al.* (Wang *et al.*, 2003) reported large room temperature saturation magnetization in epitaxial $BiFeO_3$ thin films and attributed this to thin film strain effects, subsequent reports suggested that this could be due to the presence of small amounts of $Fe^{2+}$ ions (Eerenstein *et al.*, 2005) or impurities like γ-$Fe_2O_3$ or maghemite (Bea *et al.*, 2005) which can remain undetected in the XRD pattern. It should be noted that while the measurements showed by Eerenstien *et al.* were at room temperature (Eerenstein *et al.*, 2005), the measurements showed by Bea *et al*. were conducted at 10 K (Bea *et al.*, 2005). Reports on magnetic behaviour of nanocrystalline maghemite also suggest that nanocrystalline maghemite is ferromagnetic at very low temperatures and exhibits supraparamagnetism at RT i.e. finite saturation magnetization but no remnant magnetization (Morales *et al.*, 1999). In the thin films of 300-400 nm thickness, any impurities that will be present in the samples are likely to be nanocrystalline size, also corroborated by the representative TEM micrograph as shown in Figure 1 (b) showing presence of nanocrystalline features. Hence, if at all there are any impurities like maghemite or magnetite present, these are likely to contribute only to the saturation magnetization at RT not the remanent magnetization. This again supports our argument that the presence of finite remnant magnetization in the BF-xPT multilayer samples is likely to arise either from the destruction of the spin cycloid in BF due to inhomogeneous mixing of atoms in the BF-xPT solid solution resulting in the distribution of Fe ions across the film in a manner leading to the formation of small pockets of magnetic regions of size smaller than the spin cycloid length of BF, melting the hitherto present spin spirals in homogenous BF-xPT (Zhu *et al.*, 2008).

## V. CONCLUSIONS

$(BiFeO_3)_{1-x}$-$(PbTiO_3)_x$ films fabricated by chemical solution deposition of multilayers of $BiFeO_3$ and $PbTiO_3$ containing solutions showed single phase structure with a monoclinic structure (space group Cm) as confirmed by XRD patterns and TEM images. To our surprise, we do not observe any change in the structure at around $x = 0.30$, MPB composition in the bulk, indicating disappearance of MPB, attributed to peculiar processing technique. However, broadness of XRD peaks indicates the presence of fine composition gradient across the film's cross-section resulting in unexpected as well as unusual observations in the structure evolution and the properties. Again, ferroelectric measurements do not show any particular enhancement at $x = 0.30$ and polarization of the sample marginally increased at higher BF content, an observation consistent with higher ferroelectric polarization of pure BiFeO$_3$ reported in literature. Magnetization measurements show that magnetic moment of the samples increases with increasing BF content and is attributed to the melting of the periodic spin structure. XPS measurements on the samples suggest absence of any Fe$^{2+}$ in the samples ruling out any mixed valence effect.

## ACKNOWLEDGEMENT

Authors acknowledge Defence Research and Development Organisation (Govt. of India) for the financial support for the work. AG also thanks Department of Science and Technology (India) for Ramanna Fellowship.

# References

Bea H, Bibes M, Barthelemy A, Bouzehouane K, Jacquet E, Khodan A, Contour J P, Fusil S, Wyczisk F, Forget A, Lebeugle D, Colson D and Viret M 2005 Influence of parasitic phases on the properties of $BiFeO_3$ epitaxial thin films *Applied Physics Letters* **87** 072508

Bhattacharjee S, Tripathi S and D. P 2007 Morphotropic phase boundary in $(1 - x)BiFeO_3$-$xPbTiO_3$: phase coexistence region and unusually large tetragonality *Applied Physics Letters* **91** 042903

Catalan G, Janssens A, Rispens G, Csiszar S, Seeck O, Rijnders G, Blank D H A and Noheda B 2006 Polar domains in lead titanate films under tensile strain *Physical Review Letters* **96** 127602

Eerenstein W, Morrison F D, Dho J, Blamire M G, Scott J F and Mathur N D 2005 Comment on "Epitaxial $BiFeO_3$ Multiferroic Thin Film Heterostructures" *Science* **307** 1203a

Fedulov S A, Ladyzhinskii P B, Pyatigorskaya I L and Venevtsev Y N 1964 Complete phase diagram of the $PbTiO_3$-$BiFeO_3$ system *Soviet Physics-Solid State* **6** 375

Gabbasova Z V, Kuzmin M D, Zvezdin A K, Dubenko I S, Murashov V A, Rakov D N and Krynetsky I B 1991 $Bi_{1-x}R_xFeO_3$ (R = rare-earth) - a family of novel magnetoelectrics *Phys. Lett. A* **158** 491

Gupta S, Garg A, Agrawal D C, Bhattacharjee S and Pandey D 2009 Structural changes and ferroelectric properties of $BiFeO_3$-$PbTiO_3$ thin films grown via a chemical multilayer deposition method *Journal of Applied Physics* **105** 014101

Kaczmarek W, Pajak Z and Polomska M 1975 Differential thermal analysis of phase transitions in $(Bi_{1-x}La_x)FeO_3$ solid solution *Solid State Commun* **17** 4 807

Kadomtseva A M, Zvezdin A K, Popov Y F, Pyatakov A P and Vorob'ev G P 2004 Space-time parity violation and magnetoelectric interactions in antiferromagnets *JETP Letters* **79** 571

Khan M A, Comyn T P and Bell A J 2007 Large remanent polarization in ferroelectric $BiFeO_3$-$PbTiO_3$ thin films on Pt/Si substrates *Applied Physics Letters* **91** 032901

Kumar M M, Srinivas A, Suryanarayana S V and Bhimasankaram T 1998 Dielectric and impedance studies on $BiFeO_3$-$BaTiO_3$ solid solutions *Physica Status Solidi A - Applied Research* **165** 317

Lebeugle D, Colson D, Forget A, Viret M, Bonville P, Marucco J F and Fusil S 2007 Room-temperature coexistence of large electric polarization and magnetic order in $BiFeO_3$ single crystals *Physical Review B* **76** 024116

Liu H R, Liu Z L, Liu Q and Yao K L 2006 Electric and magnetic properties of multiferroic $(BiFeO3)_{(1-x)}$-$(PbTiO3)_{(x)}$ films prepared by the sol-gel process *Journal of Physics D-Applied Physics* **39** 1022

Morales M P, Veintemillas-Verdaguer S, Montero M I, Serna C J, Roig A, Casas L, Martinez B and Sandiumenge F 1999 Surface and internal spin canting in gamma-$Fe_2O_3$ nanoparticles *Chem Mater* **11** 3058

Sakamoto W, Yamazaki H, Iwata A, Shimura T and Yogo T 2006 Synthesis and characterization of $BiFeO_3$-$PbTiO_3$ thin films through metalorganic precursor solution *Japanese Journal of Applied Physics Part 1-Regular Papers Brief Communications & Review Papers* **45** 7315

Schedel-Niedrig T, Weiss W and Schlögl R 1995 Electronic structure of ultrathin ordered iron oxide films grown onto Pt(111) *Phys Rev B* **52** 17449

Sunder V, Halliyal A and Umarji A M 1995 Investigation of tetragonal distortion in the PbTiO$_3$-BiFeO$_3$ system by high-temperature x-ray diffraction *Journal of Materials Research* **10** 1301

Vanderbilt D and Cohen M H 2001 Monoclinic and triclinic phases in higher-order Devonshire theory *Physical Review B* **63** 094108

Wang J, Neaton J B, Zheng H, Nagarajan V, Ogale S B, Liu B, Viehland D, Vaithyanathan V, Schlom D G, Waghmare U V, Spaldin N A, Rabe K M, Wuttig M and Ramesh R 2003 Epitaxial BiFeO$_3$ multiferroic thin film heterostructures *Science* **299** 1719

Zhu W-M, Guo H-Y and Ye Z-G 2008 Structural and magnetic characterization of multiferroic (BiFeO$_3$)$_{1-x}$(PbTiO$_3$)$_x$ solid solutions *Physical Review B* **78** 014401

**List of figures**

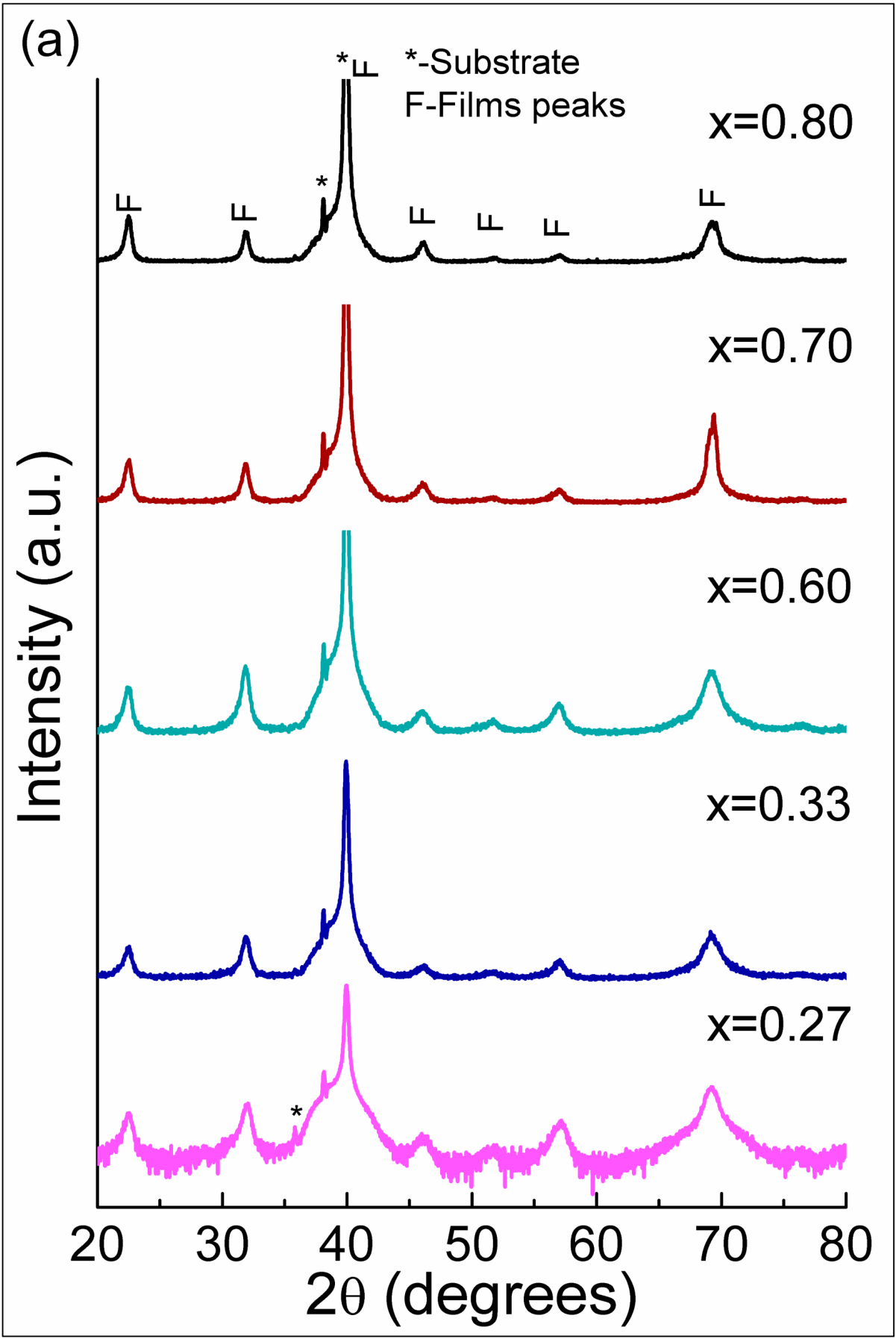

**Figure 1a**

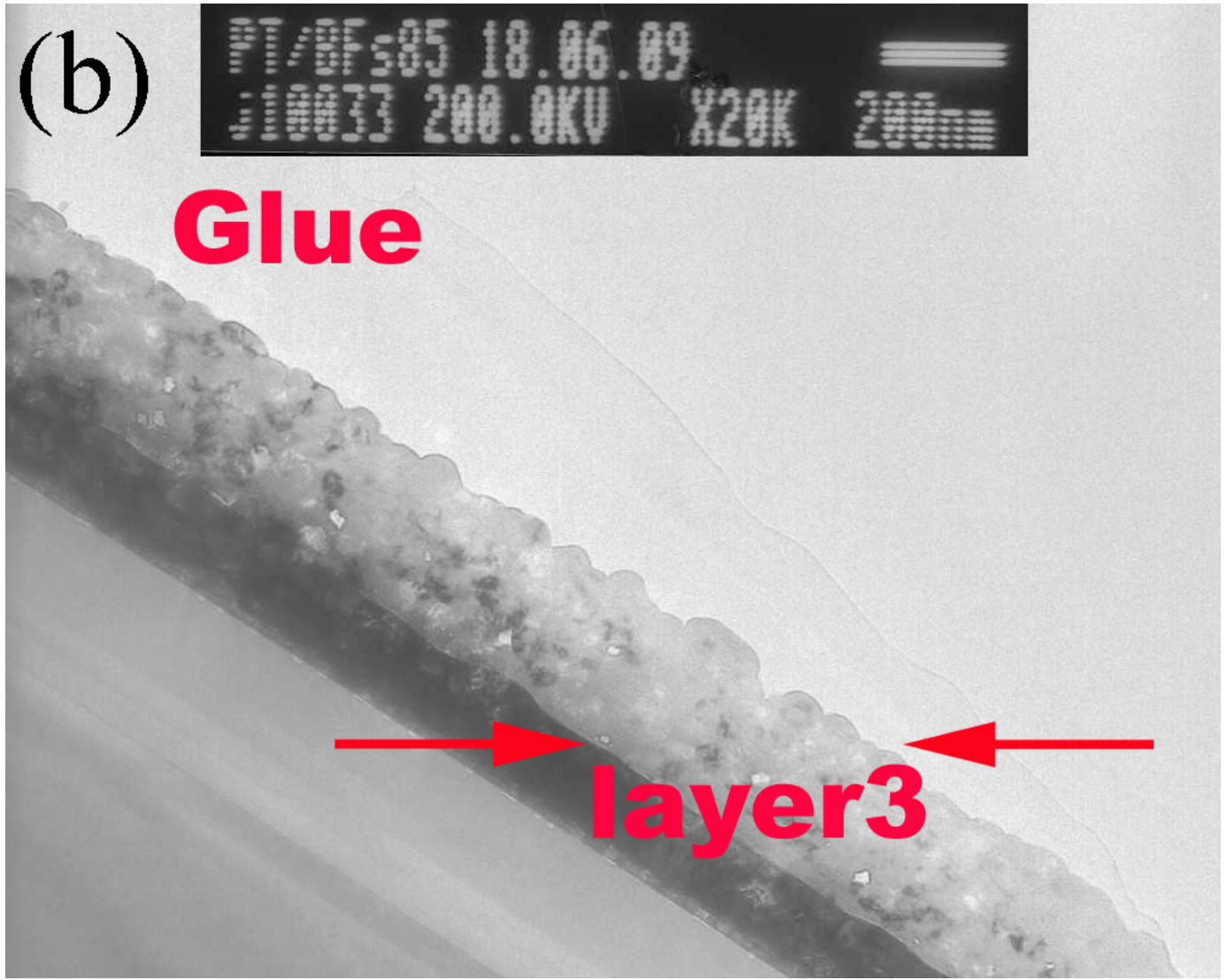

Figure 1b

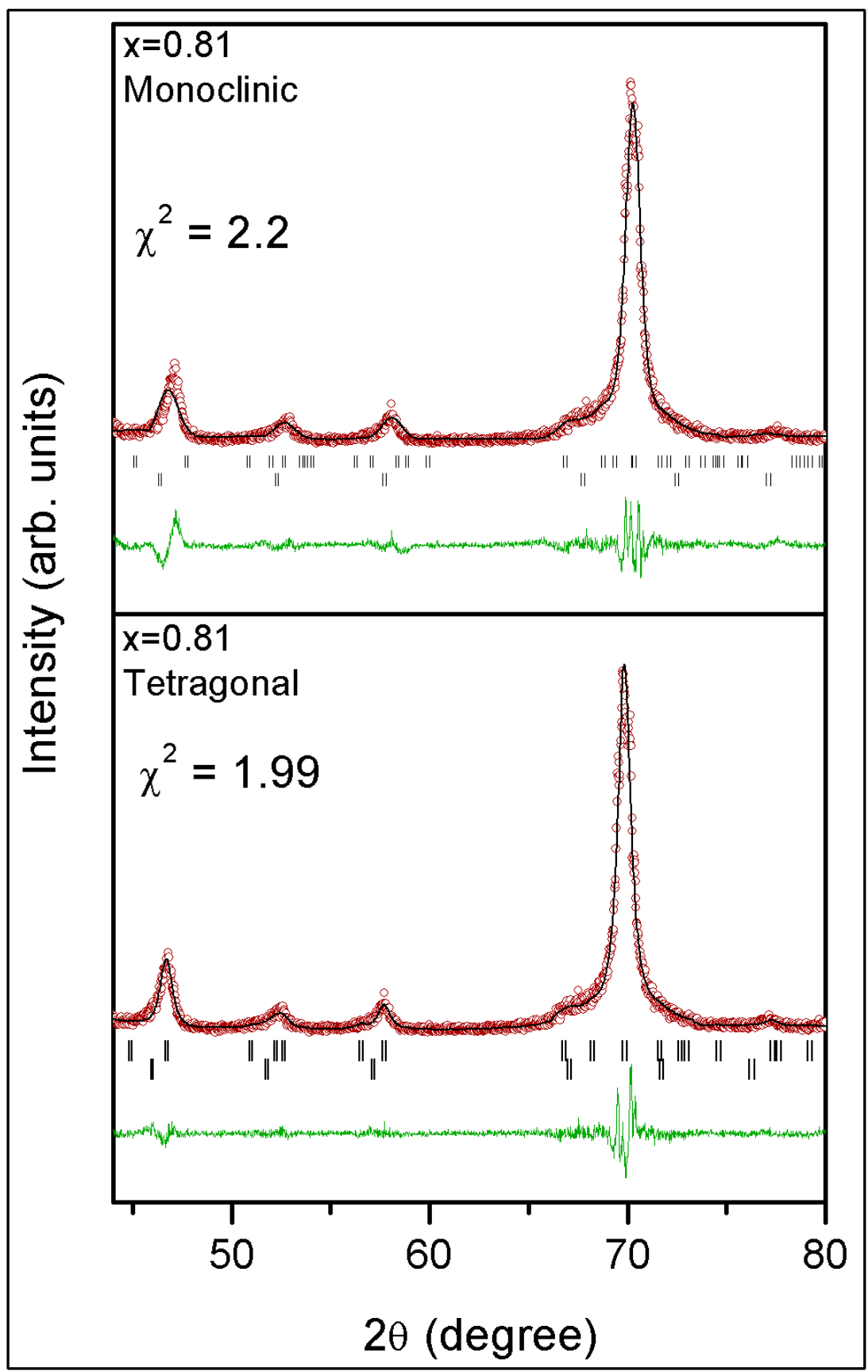

Fig. 2

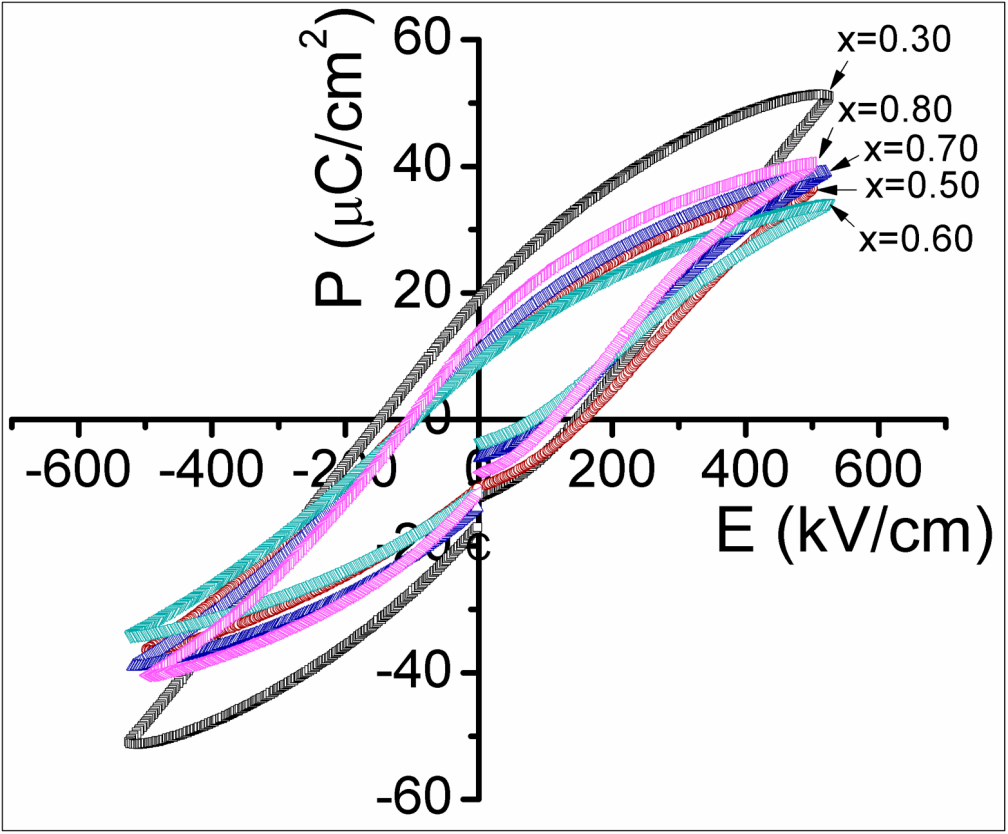

Figure 3

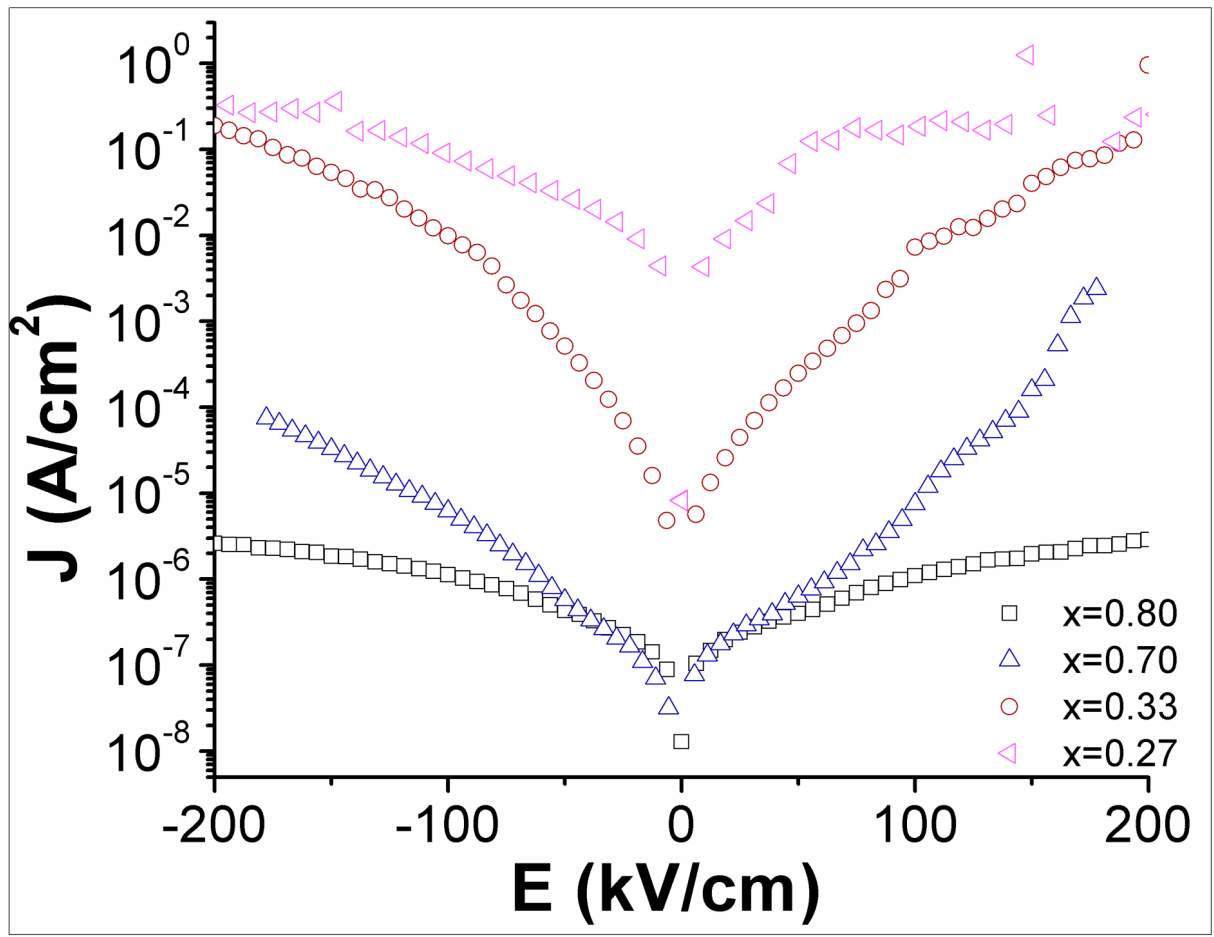

Figure 4

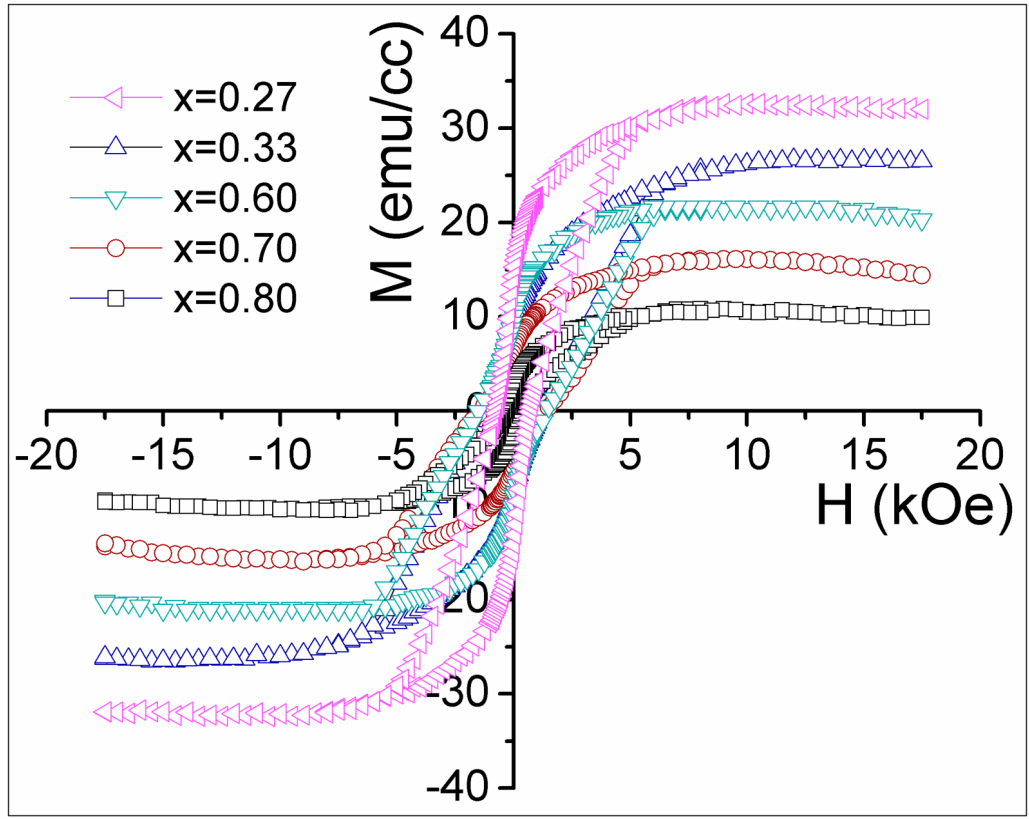

Figure 5

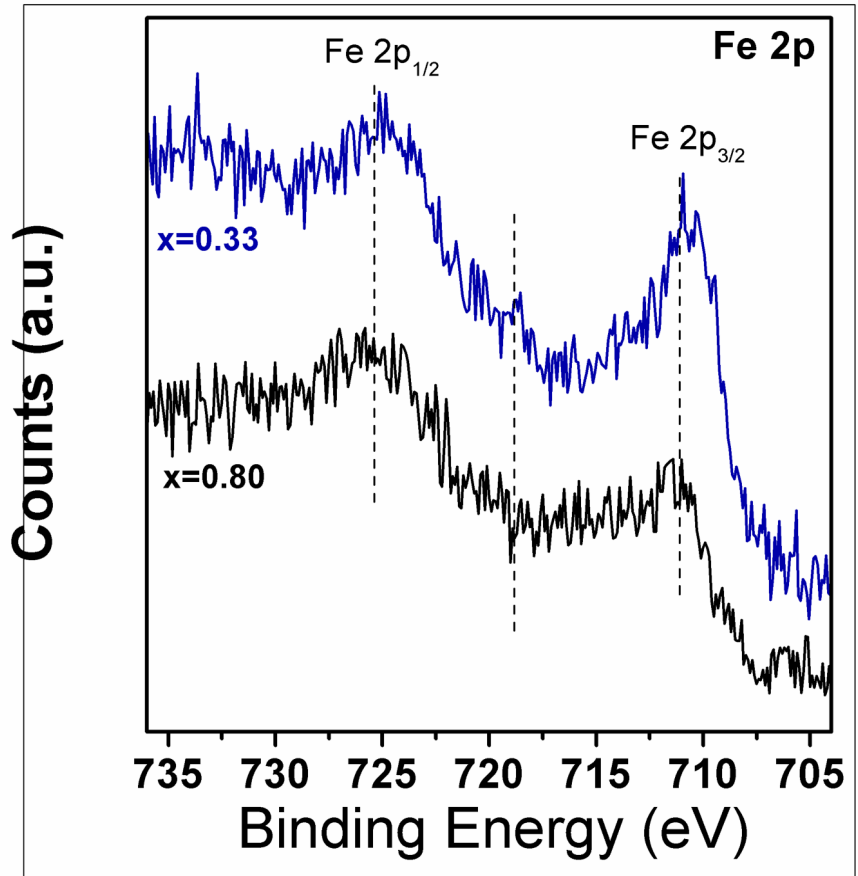

Figure 6 (a)

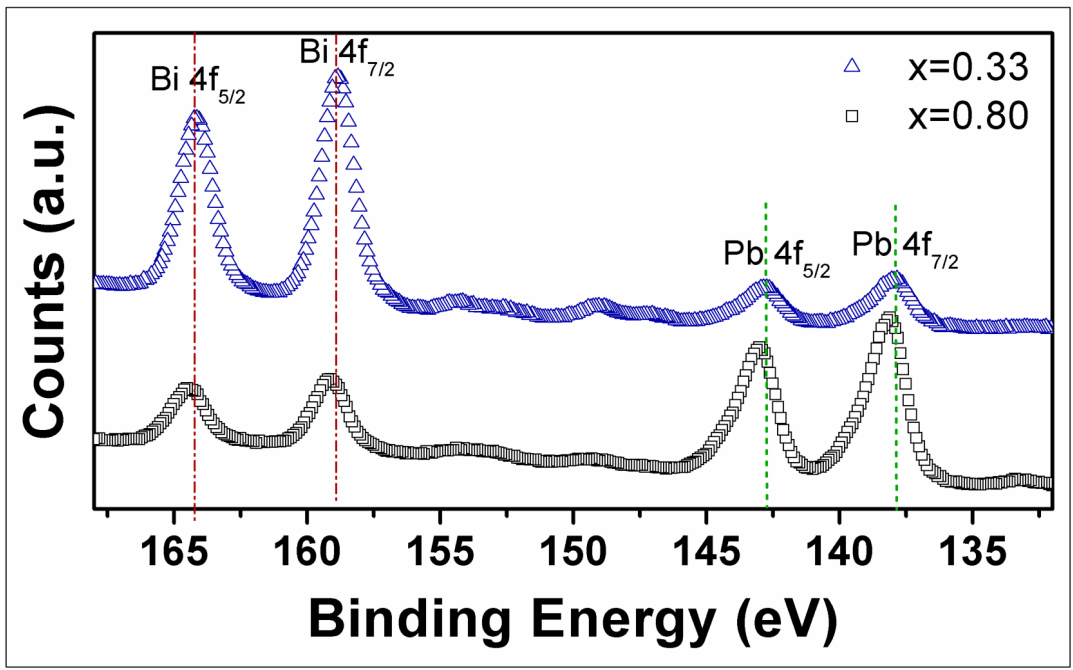

Figure 6 (b)